\documentclass[showpacs,twocolumn,prb,aps]{revtex4}
\usepackage{epsfig}
\usepackage{amsmath}
\usepackage{amssymb}
\begin{document}
\title{Spin Conductance in one-dimensional Spin-Phonon systems}

\author{Kim Louis and Xenophon Zotos} 

\affiliation{
Department of Physics, University of Crete and Foundation
for Research and Technology-Hellas, P. O. Box 2208, 71003 Heraklion,
Crete, Greece
}

\date{\today}

\begin{abstract}
We present results for the spin conductance of the one dimensional 
spin-1/2 Heisenberg  and  $XY$ model
 coupled to phonons. We apply an approach 
based on the Stochastic Series Expansion (Quantum Monte Carlo) method 
to evaluate the conductance for a variety of phonon dispersions 
and values of spin-phonon coupling. 
From our numerical simulations and analytical arguments,
we derive several scaling laws for the conductance.

\end{abstract}
\maketitle

{\it Introduction} ---
Controlling and understanding the dynamics and transport of spins 
is an experimental and theoretical challenge aiming at the development of 
mesoscopic systems for spintronics applications.
In particular, the interest is recently focusing on the effect of 
interactions and the coupling of spin systems to the environment;  
several proposals for the generation and detection of spin 
currents exist. \cite{spintronics}
As a key quantity characterizing 
ballistic mesoscopic spin transport is the (spin) conductance, it is 
important to understand its behavior in prototype interacting spin systems 
coupled to lattice vibrations -- phonons.

Related to this problem, since the 60's, the spin-Peierls transition, 
the dimerization of a lattice due to the spin-phonon interaction in the bulk, 
has been discussed for quasi-one dimensional compounds as TTF, TCNQ, 
CuGeO$_3$. A variety of analytical and numerical methods are still employed in 
the study of the dependence of the transition on the 
phonon spectrum and spin-phonon coupling. \cite{cross,Sandvik1} 

Finally, in close relation to this work, 
spurred by the experimental finding of an exceptionally
high thermal conductivity \cite{sol01,hes01} in quasi-one dimensional 
transition metal compounds, 
there is recently considerable experimental and theoretical research interest 
in the finite temperature transport properties of spin 
models. \cite{review,Alv02,HM} In particular, this activity leads to the 
development of novel 
numerical simulation techniques \cite{techpap} for the calculation of 
zero and finite temperature transport properties in one dimensional 
electronic/spin models.

In this paper we address the fundamental question of the 
spin conductance of the spin-1/2 Heisenberg model and the $XY$ model
 coupled to phonons. 
Thus this study, on the one hand, sheds light from a new perspective on 
the old question 
of the spin-Peierls transition and on the other hand addresses a prototype 
question in mesoscopic physics;  
it is of direct experimental relevance, as a detailed
experimental set-up for the observation of the spin conductance
 was recently proposed. \cite{MeiLos}

Quantum Monte Carlo (QMC) methods  have
 already been successfully applied to the 
evaluation of thermodynamic quantities in spin-phonon
systems. \cite{Sandvik1,Sandvik2,Loew}
We will use here the Stochastic Series Expansion (SSE)  which allows
the calculation of
 the conductance of interacting 
spin systems; \cite{techpap} in this original application we extend it to 
coupled spin-phonon systems.

In the following we will first discuss the model and the method we are using,
then  we will present data for the conductance in the Heisenberg model
for optical and acoustic phonon dispersions, and derive several  
scaling laws. 
Finally, we will show results for the $XY$ model coupled to optical phonons. 

{\it Model and method} --- The nominal spin-Peierls Hamiltonian 
describing the coupling of the one dimensional Heisenberg model 
to lattice vibrations is given by
\begin{eqnarray}
H&=&\sum_n\left[J+\lambda(x_n-x_{n+1})\right]({\bf S}_n\cdot 
{\bf S}_{n+1})\nonumber \\
&+&\sum_n\omega_0/2\left[2{p_n^2}+(x_n-x_{n+1})^2\right],
\label{HI}
\end{eqnarray}

\noindent
where ${\bf S}_n$ are spin-1/2 operators and $x_n$, $p_n$ 
the displacement and momentum  operators of the ions 
($\omega_0$ is the characteristic phonon frequency and here we set
$\hbar=1$). 

Because of the minus sign problem, this system eludes a direct 
analysis using the QMC method that we intend 
to employ here. Hence, we resort to  the following generalized 
Hamiltonian which is an extension of the Hamiltonian discussed 
in Refs.\ \onlinecite{Sandvik1,Sandvik2}, 

\begin{eqnarray}
H&=&\sum_{n=0}^{N-1}\left(\tilde J+\alpha_1(a_n+a_n^\dag)+
\alpha_2(a_{n+1}+a_{n+1}^\dag)\right)\cdot h_n\nonumber \\
&+&\sum_{n=0}^{N-1}\Bigl[\omega (a_n^\dag a_n^{\phantom{\dag}}+a_{n+1}^\dag 
a_{n+1}^{\phantom{\dag}})-\omega_1(a_n^\dag a_{n+1}+a_{n+1}^\dag a_{n})
\nonumber \\
&&\phantom{\sum}-\omega^+
(a_n^\dag a_{n+1}^\dag+a_{n+1}
a_{n})\Bigr],
\label{HII}
\end{eqnarray}
where $h_n= \tilde C+S_n^zS_{n+1}^z-(S_n^+S_{n+1}^-+S_n^-S_{n+1}^+)/2$ and 
$a_n$ ($a_n^{\dag}$) are phonon annihilation (creation) 
operators. From Eq.\ (\ref{HII}) one recovers the spin-Peierls 
model Eq.\ (\ref{HI}) if one sets

\begin{equation}
\lambda/\sqrt{2}=\alpha_1=-\alpha_2,\qquad 
\omega_0=\omega=2\omega_1=2\omega^+
\label{ItoII}
\end{equation} 

\noindent
and $\tilde C=0$, $\tilde J=J$. 
However, as a finite $\tilde C<-1/4$ is required for the implementation of 
the SSE \cite{Sandvik,SandSyl} we shift (following Ref.\ \onlinecite{Sandvik1}) 
the boson operators $a_n$ and $\tilde J$,

\begin{eqnarray} 
a_n&\to& a_n-(C/2)(\alpha_1+\alpha_2)/
(\omega-\omega_1-\omega^+)\label{replace1}\\
\tilde J&\to& \tilde J+(\alpha_1+\alpha_2)^2C/(\omega-\omega_1-\omega^+)
\label{replace2}
\end{eqnarray}

\noindent
so that $\tilde C=C,\tilde J\neq J$.
Still, to sample the Hamiltonian Eq.\ (\ref{HII})
with the SSE we have to meet the
requirement that all matrix elements of $H$ are  
negative, \cite{Sandvik,SandSyl} implying
 $\alpha_1>0, \alpha_2>0, \tilde J>0$, or explicitly ($C=-1/4$):
\begin{equation} 
(\alpha_1+\alpha_2)^2/4\leq \omega-\omega_1-\omega^+,
\qquad \alpha_1,\alpha_2\geq 0.
\label{cond}
\end{equation}

\noindent
The conditions in Eq. (\ref{cond}) thus forbid a direct analysis of the
spin-Peierls Hamiltonian Eq.\ (\ref{HI}) as may be seen from Eq.\ 
(\ref{ItoII}). However, as we will see we find two scaling laws
which enable us to make predictions for the spin-Peierls model Eq. (\ref{HI}).

We define the operators $j_n$ and $P_m$,  
$$j_n=\frac{iJ_xe}{2\hbar}\left(S_n^+S_{n+1}^--S_n^-S_{n+1}^+\right)\qquad
P_m=e\sum_{n>m}S^z_n, $$
which are the current operator for magnetic moments at site $n$ 
(we denote the ``charge'' quantum by $e$  in analogy to the electric current) 
and  the potential operator of
a local ``voltage drop'' at site $m$, respectively.
The spin conductance $g$ is the linear response of the operator $j_n$
to the perturbation $P_m$, 
$$
g:=\lim_{z\to 0}{\rm Re}\int_0^\infty e^{izt}\frac{i}{\hbar} \langle
[j_x(t),P_y]\rangle \,dt.
$$
It is independent  of $n$ or $m$ and will be evaluated
according to Ref.\ \onlinecite{techpap} 
while the SSE method for the Hamiltonian Eq.\ (\ref{HII}) 
is implemented following  Refs.\ \onlinecite{Sandvik1,Sandvik2}. 
Note that we are considering only low temperatures and the 
effect of leads on the conductance is not included.

For the update of the spin degrees of freedom we use the very efficient
scheme of ``operator-loop updates''. \cite{SandSyl}  
The simplest version of an update for the phonons is to suggest a specific 
change in the configuration and accept/reject it with 
a certain probability. Here we want to emphasize that we obtain smaller 
autocorrelation times by using an appropriate cluster update
scheme, reminiscent of the loop update (see appendix, section \ref{update}).

In our simulations we use a system size $N=192$, a temperature
$T=0.02 J/k_B$ and typically perform $2\cdot 10^5$ MC sweeps.

It should also be mentioned that in our simulations 
there is no need to introduce a bound for the phonon occupation number, 
but for technical reasons we impose a restriction of maximum $48$ phonons
per site; 
since it is much larger than the phonon numbers created 
during the simulation this introduces no further error.

{\it Optical Phonons} --- 
The two parameters $\omega_1$ and $\omega^+$ lead to 
a nontrivial dispersion relation of the phonon modes.
For the moment we will set them to zero and consider
phonons of constant frequency $\omega$.
This (Einstein) model was discussed 
for $\alpha_1=-\alpha_2$ using the DMRG method \cite{Bursill}  
and for $\alpha_2=0$ with the SSE. \cite{Sandvik2}
Both find that if $\alpha_1$ is larger than a 
critical coupling $\alpha_c$ the system 
undergoes a dimerization transition.  In Ref.\ \onlinecite{Bursill}
the authors analyze different values for $\omega$ finding that $\alpha_c$ 
is approximately linear in $\omega$ if $\omega<J$.
The authors in Ref.\ \onlinecite{Sandvik2} discuss only one phonon frequency,
namely $\omega=J/8$. They find that $\alpha_c=1.3\omega$. (Data in our notation.) We will use this  result since we focus on $\omega < J$.
 The reason for this is that in many materials of 
actual interest $J$ is significantly larger than the Debye temperature;  
we also find that our QMC method converges better for small $\omega$.

\begin{figure}[t]
\epsfig{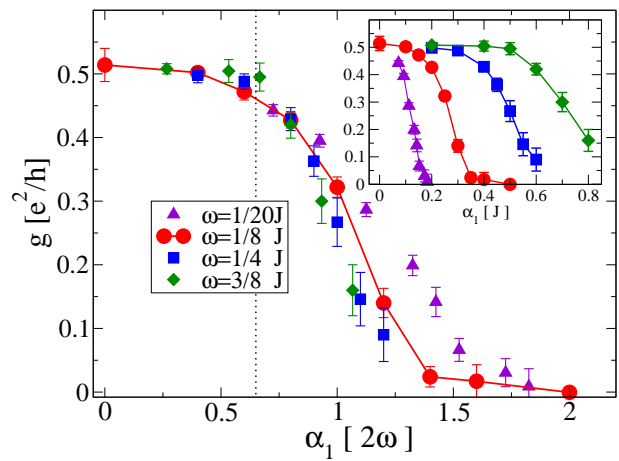}
\caption{The conductance as a 
function of $\alpha_1$ for various $\omega<J$
for optical phonons  ($\omega_1=0=\omega^+$ and 
$\alpha_2=0$). $\alpha_1$ is given in units $2\omega$.
The dotted line is situated at the critical coupling
 of Ref.\ \onlinecite{Sandvik2}.
 The inset shows 
the same curves with $\alpha_1$ in units of $J$.}
\label{gphowi}
\end{figure}

For the conductance $g(T=0)$ we expect the form of a step function: Above the critical 
coupling $\alpha_1>\alpha_c$ it is zero due to the opening of the gap;
for $\alpha_1<\alpha_c$ the  system is supposed to be in a Luttinger phase,
where the conductance is given by the Luttinger parameter $K$. 
Since the Hamiltonian Eq.\ (\ref{HII}) is spin-rotationally invariant, 
the latter is fixed  to the value of $1/2$ (see Ref.\ \onlinecite{Schulz}).

The fact that $g$ should be constant for $\alpha_1 \leq \alpha_c$
 is consistent 
with an effective Hamiltonian that we can derive using 
second order perturbation theory (cf.\  Refs.\ \onlinecite{Bursill,perturb})

\begin{equation}H=\sum_n
\left(1+\frac{\alpha_1^2+\alpha_2^2}{4\omega}\right) 
{\bf S}_n\cdot {\bf S}_{n+1}-
\frac{\alpha_1\alpha_2}{4\omega} {\bf S}_n\cdot {\bf S}_{n+2}\, .
\label{Hper}
\end{equation}
Indeed, for $\alpha_2=0$ we find no change in the conductance to order 
$\alpha_1^2$ since $g$ does not depend on the bandwidth $J$.

In Fig.\ \ref{gphowi}, we present results for the spin conductance $g$ 
 for $\omega_1=0=\omega^+$ and $\alpha_2=0$. 
Within error bars we confirm the expectation that the conductance remains
at  $1/2$ until the critical coupling is reached.  In the gapped phase
 the conductance goes rather smoothly to zero instead of step-like which is
 due to the
finite temperature $T=0.02J$ we are using.  The point where
the conductance has decayed to zero  gives a rough estimate for 
the coupling $\alpha_T$ where
the gap satisfies $\Delta(\alpha_T)=k_BT$.  
(This can be tested  using data from Ref.\ \onlinecite{Bursill}, if
we assume that $\Delta(\alpha_1,\alpha_2)=\Delta(\alpha_1-\alpha_2)$---which is a natural assumption since this is the scaling relation of $\alpha_c$ to be derived below, cf.\ Eq.\ (\ref{scale3}).
For $\omega=J/20$ we find $\alpha_{T=0.02J}-\alpha_c\approx 0.088J$, i.e.,
$(\alpha_{T=0.02J}-\alpha_c)/\omega^{0.73}\approx 0.784$ to be compared
 with Fig.\ \ref{gphocc3} below.)

Subsequent calculations we performed  for  $T=0.01J$
 find an increased statistical
error, but reveal no temperature dependence which implies that
 $\alpha_{T=0.01J}$ and $\alpha_{T=0.02J}$
are very close to each other. Note that the gap has a typical 
Kosterlitz-Thouless form:\cite{Bursill}
 first it opens very slowly, but then it grows rather rapidly.

Since $g(T=0)$ is a step function---which is fully described by the critical
coupling where the conductance drops to zero---it  inherits the scaling 
law of $\alpha_c$:
\begin{equation}  
g(\alpha_1,\omega)=g(\alpha_1/\omega) \quad {\rm for}\;T=0.
\label{scale1}
\end{equation}  

\noindent

\begin{figure}[h]
\epsfig{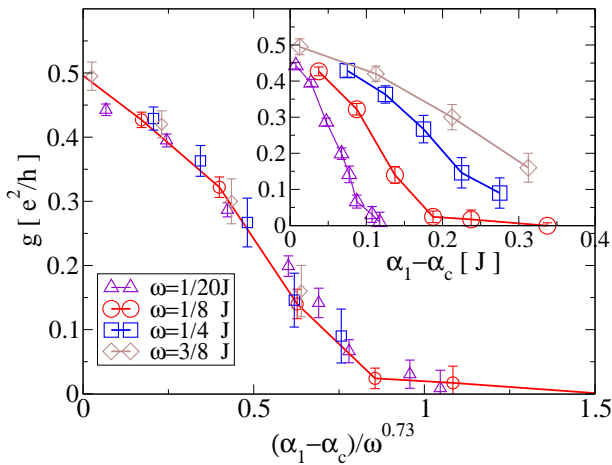}
\caption{
The conductance as a function of $(\alpha_1-\alpha_c)/\omega^{0.73}$.
The inset shows the same data as a function of $\alpha_1-\alpha_c$ 
in units of $J$ .
}
\label{gphocc3}\end{figure}

To analyze the behavior of $g$ in the gapped region more closely,
 it is instructive to plot
$g$ versus $\alpha_1-\alpha_c$. This is done in Fig.\ \ref{gphocc3}.
We find good agreement with another scaling relation---different from Eq.\ (\ref{scale1})---specific to the gapped phase, namely,
\begin{equation}g(\alpha_1,\omega)=g\left[(\alpha_1-\alpha_c)/\omega^\kappa\right]
\label{scale2g}.\end{equation}
where $\kappa=0.73\pm 0.05$. It is natural to assume that
in the gapped region $g$ depends on the parameters of the system only through
 $\beta\Delta$.
This would in turn imply that the scaling relation Eq.\ (\ref{scale2g})
 holds for the gap $\Delta$. Note that this relation is not consistent with the
exponential behavior of $\Delta$ typical for a Kosterlitz-Thouless transition.
However, we should expect Eq.\ (\ref{scale2g}) to be approximately valid for 
not too large $\alpha$ and $\omega$. We note that
the data for $\Delta(\omega=0.05,0.5)$ 
from Ref.\ \onlinecite{Bursill} show also good agreement with Eq.\ (\ref{scale2g})
 when $\kappa=0.6$.

\begin{figure}[h]
\epsfig{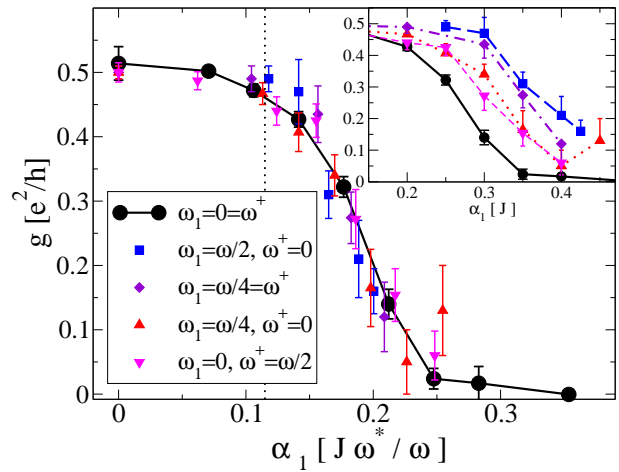}
\caption{
The conductance as a function of $\alpha_1$ (in units of $J\omega^\ast/\omega$)
($\alpha_2=0,\; \omega=J/8$) for various dispersions $\omega (k)$.
The dotted line indicates the critical coupling
 of Ref.\ \onlinecite{Sandvik2}.
The inset shows the same data when $\alpha_1$ is given in units of $J$.
}
\label{akpho2}\end{figure}

{\it Acoustical phonons} ---
Now we will consider nontrivial phonon dispersions, 
$\omega_1\neq0\neq\omega^+$. Performing a canonical Bogoliubov 
transformation in momentum (k-) space, we obtain the 
phonon frequency dispersion

\begin{equation}
\omega(k)=2\sqrt{[\omega-\omega_1\cos(k)]^2
-[\omega^ +\cos(k)]^2}\, .
\label{omegk}
\end{equation}

\begin{figure}[h]
\epsfig{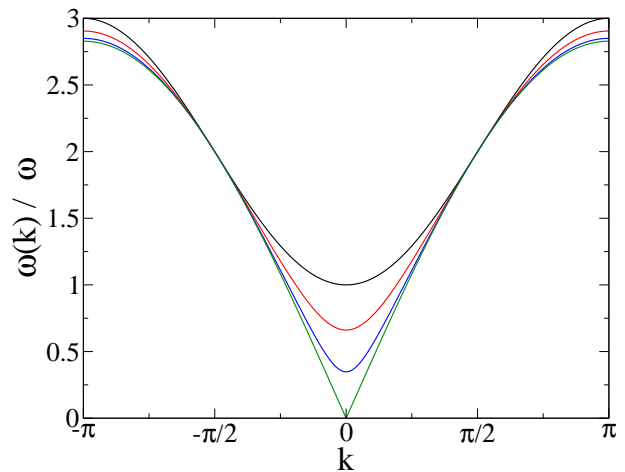}
\caption{The phonon dispersion Eq.\ (\ref{omegk}) for $\omega_1=\omega/2$
and $\omega^+$ is $0$, $(3/4)\omega_1$, $(15/16)\omega_1$, 
$\omega_1$ (from top to bottom).
At $\omega^+=0$ the parameter  $\omega_1$ gives the width of the dispersion
and $\omega$ its mean value.}
\label{omphonk}\end{figure}

We see that if  $\omega=\omega_1+\omega^+$,  as, e.g. in the spin-Peierls
model Eq.\ (\ref{HI}), $\omega(k=0)=0$, whereas if 
$\omega> \omega_1+\omega^+$ there is a gap in the spectrum. 
This dispersion is illustrated in Fig.\ \ref{omphonk}.

After diagonalizing the phonon part of the Hamiltonian Eq.\ (\ref{HII})
we proceed similarly with the spin-phonon coupling term.
To this end, adopting for convenience a fermion representation 
(${\bf S}^+ \mapsto c_n^{\dagger}$) by the Jordan and Wigner 
transformation, the phonon-fermion scattering term 
$$\frac{\alpha_1}{2}\sum_nc_n^\dag c_{n+1}(a_n+a_{n}^\dag)$$
is written in k-space as
$$\frac{\alpha_1}{2}\sum_{k,q}\frac{\sqrt{\omega(k-q)/N}\exp(-iq)}
{\sqrt{2\omega-2\cos(k-q)[\omega_1+\omega^+]}}c_k^\dag c_{q}
(a_{k-q}+a_{q-k}^\dag).$$

Since we are here interested only in the low temperature regime 
we may assume that the  variables $k,q$ lie at the Fermi momenta $\pm \pi/2$.
The process responsible for the decrease of the 
conductance should then be the 
scattering between two Fermi points by a phonon with momentum $\pi$ and 
the only relevant frequency should be $\omega(\pi)$.
Thus, for a general dispersion relation, 
we should recover the same result for $g$ as for the optical phonons of the 
previous section, by mapping $ \omega \to \omega(\pi)/2$,  
$\alpha_1 \to \alpha_1\sqrt{\omega(\pi)}/\sqrt{2(\omega+\omega_1+\omega^+)}$. 
We expect then the conductance $g$ to depend only on $\alpha_1/\omega^\ast$ 
with $\omega^\ast=\sqrt{(\omega+\omega_1+\omega^+)\omega(\pi)}$;
to verify this expectation we show $g$ in Fig.\ \ref{akpho2} for a set 
of parameters $\omega$, $\omega_1$, $\omega^+$ 
compatible with Eq. (\ref{cond}).

The figure indeed suggests that 
$g(\alpha_1,\omega,\omega_1,\omega^+)=g(\alpha_1\omega/\omega^\ast,\omega)$ 
and together with Eq. (\ref{scale1}) we arrive at the desired 
scaling relation

\begin{equation}
g(\alpha_1,\omega,\omega_1,\omega^+)=g(\sqrt{2}\alpha_1/\omega^\ast),\quad {\rm for}\;
T=0.
\label{scale2}
\end{equation}

\begin{figure}[h]
\epsfig{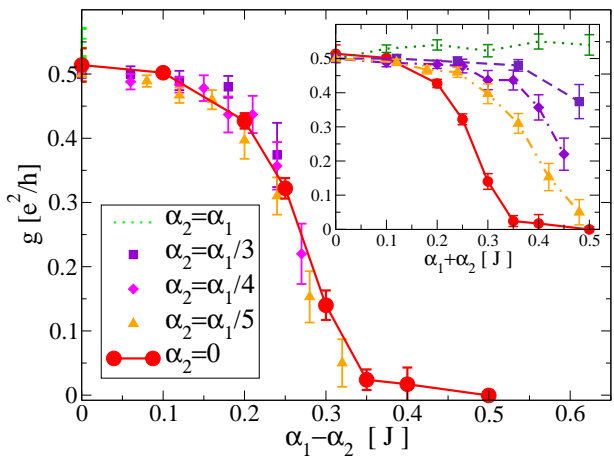}
\caption{The conductance as a 
function of $\alpha_1-\alpha_2$ for various  $\alpha_2/\alpha_1$.
(The inset shows $g$ vs $\alpha_1+\alpha_2$. The order of the curve
-top to bottom- is the same as in the legend.)
Here $\omega=J/8$, $\omega_1=0=\omega^+$, $T=0.02J/k_B$.
}
\label{gaaic}\end{figure}

{\it Scaling with $\alpha_1$ and $\alpha_2$} ---
Let us again consider the Hamiltonian in k-space as in the previous section.
If we have $\alpha_1$ and $\alpha_2$ different from zero,
 the coupling terms read
\begin{equation}\frac{1}{2}\sum_{k,q}A_{k,q}
(\alpha_1+e^{i(k-q)}\alpha_2)c_k^\dag c_{q}
(a_{k-q}+a_{q-k}^\dag).
\label{lowenergy}\end{equation}
Here $A_{k,q}$ is a prefactor---whose precise form is for the following discussion irrelevant.
At low temperatures we have again $k,q=\pm \pi/2$ which implies that
 $k-q=0,\pi$. Hence only two phonon modes are important: $\omega(0)$ and
 $\omega(\pi)$. The $\omega(0)$ mode describes scattering on one Fermi point  which means that
it probably rescales the parameters of the system, but   has otherwise no
 influence on the transition. 
For the phase transition only the $\omega(\pi)$ mode  is important.
As we see from Eq.\ (\ref{lowenergy})
the  mode with $\omega(0)$ couples to the spin degrees of freedom
via $\alpha_1+\alpha_2$ and the mode $\omega(\pi)$ via $\alpha_1-\alpha_2$.
As the decay of the conductance is due to the opening of the gap, 
 we are led to assume that $g$ depends on $\alpha_1-\alpha_2$, but not on 
$\alpha_1+\alpha_2$, i.e.,

\begin{equation} g(\alpha_1,\alpha_2)=g(\alpha_1-\alpha_2).\quad{\rm for}\; T=0
\label{scale3}
\end{equation}

(Alternatively, one can motivate this scaling law with a Mean Field Theory ansatz which is outlined in the  appendix, section \ref{MeanField}.)

In Fig.\ \ref{gaaic} we confirm this scaling by the overlap of 
curves as a function of $\alpha_1 - \alpha_2$, 
while in the inset we indeed find that the 
dimerization transition is delayed
by increasing $\alpha_2/\alpha_1$, till $\alpha_2=\alpha_1$
where it is absent as predicted by Eq.\ (\ref{scale3}).

A further test of Eq.\ (\ref{scale3}) is provided by comparing data
for $\alpha_c$. For $\omega=J/8$ and $\alpha_2=0$ Ref.\ \onlinecite{Sandvik2}
  finds 
$\alpha_c=0.163J$ (with $4\%$ uncertainty), whereas according to Ref.\ \onlinecite{Bursill} for the same $\omega$
but $\alpha_1=-\alpha_2$ the transition occurs when $\alpha_1-\alpha_2=0.176J$.
The difference of the two values is $8\%$ which means reasonable
agreement.

\begin{figure}[h]
\epsfig{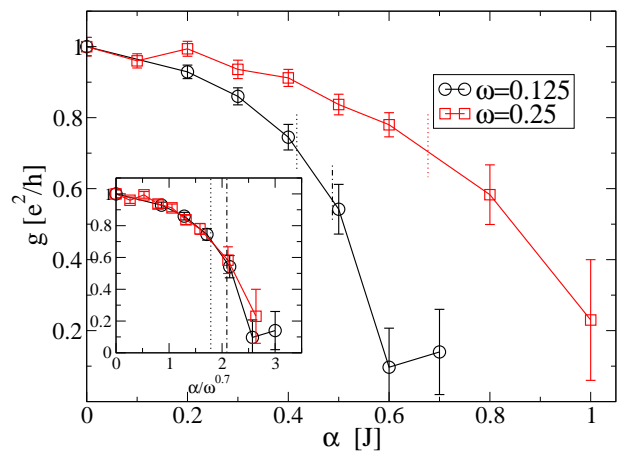}
\caption{The conductance of the $XY$ model as a 
function of $\alpha_1$ ($\alpha_2=0$) for two $\omega$ (here $T=0.01J$).
The dotted line gives the critical coupling if one assumes that the scaling law $\alpha_c(\alpha_1,\alpha_2)=\alpha_c(\alpha_1-\alpha_2)$ holds. The dot-dashed line gives the critical $\alpha_c$ extracted from Fig.\ \ref{gapJ0}.}
\label{gpho0}
\end{figure}
\begin{figure}[h]
\epsfig{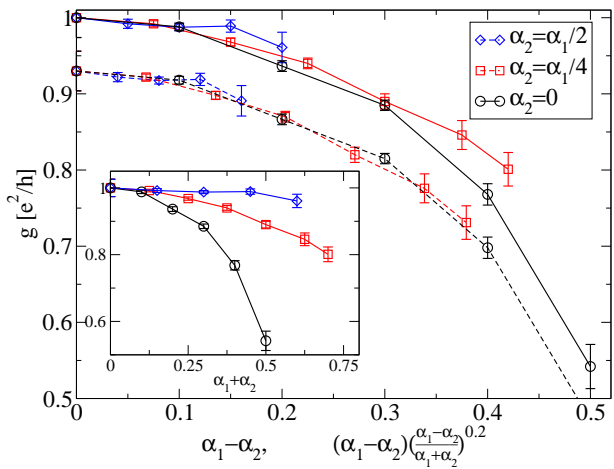}
\caption{The conductance of the $XY$ model as a 
function of $\alpha_1-\alpha_2$ (solid curves), $(\alpha_1-\alpha_2)^{1.2}/
(\alpha_1+\alpha_2)^{0.2}$ (dashed curves which are offset by $-0.07$ for clarity) and $\alpha_1+\alpha_2$ (inset) for various $\alpha_2/\alpha_1$ with $\omega=J/8, T=0.01J$. }
\label{gph0c}
\end{figure}

{\it $XY$ model}--- In this section we turn to the $XY$ model,
 that is we replace
${\bf S}_n\cdot {\bf S}_{n+1}$  in the Hamiltonian Eq.\ (\ref{HII}) by
$S_n^xS_{n+1}^x+S_n^yS_{n+1}^y$. 
The model was studied by Caron and Moukouri \cite{Caron96} for $\alpha_1=-\alpha_2$.
They found a phase transition similar to the one in the spin-Peierls
model but here $\alpha_c\propto \omega^{0.7}$. We computed the conductance
for two phonon frequencies. The results are shown in Fig.\
\ref{gpho0}. We find that the zero temperature conductance
obeys the scaling law
\begin{equation}g(\alpha_1,\omega)=g(\alpha_1/\omega^{0.7}).
\label{scale0}\end{equation}
It is important to point out that the $XY$ model  is not spin-rotationally
 invariant
and therefore the value of the conductance is not fixed in the Luttinger phase,
but rather decreases monotonously, as the Fig.\ \ref{gpho0} shows.
In the gapped phase the conductance decays again quite smoothly to zero,
 such that it is barely possible 
to decide  where exactly the transition takes place.

We therefore employed  here two different methods for calculating $\alpha_c$.
 First  QMC simulations for the staggered kinetic energy which give a rigorous upper bound (details are deferred to the appendix, section \ref{critical}), second we used the data for $\alpha_c$ from 
Ref.\ \onlinecite{Caron96} while assuming that the scaling law 
Eq.\ (\ref{scale3}), $g(\alpha_1,\alpha_2)=g(\alpha_1-\alpha_2)$, holds

 The two  results for $\alpha_c$ are 
indicated by lines in Fig.\ \ref{gpho0} for comparison. 
The difference between the two predictions for $\alpha_c$ is about $15\%$.
One should emphasize again that both methods are not numerically exact:
the QMC yields 
only an upper bound for $\alpha_c$, 
and the other method relies  on the validity of Eq.\ (\ref{scale3})
which is examined  in Fig.\ \ref{gph0c} by considering nonzero $\alpha_2$. 
For the parameter accessible to us Eq.\ (\ref{scale3}) seems to be
 approximately, but {\em not exactly}  fulfilled.

 In the Heisenberg model the dependence on $\alpha_1,\alpha_2$ and $\omega$
was described by two separate scaling laws Eqs.\ (\ref{scale1}) and
(\ref{scale3}). For the $XY$ model the situation is more complicated,
as we found that for $\alpha_2=\alpha_1/4$ the scaling law 
Eq.\ (\ref{scale0}) is violated.  This implies that a generalization of
Eq.\ (\ref{scale3}) which holds for the $XY$ model must depend on $\omega$.
We will not try to find such a generalization here, but want
to point out that for the special case of $\omega=J/8$  one can obtain 
a much better fit for our data (for $\alpha_2/\alpha_1=0,0.25,0.5$) 
than Eq.\ (\ref{scale3}) by using the following modified
 scaling ansatz (see Fig.\ \ref{gph0c})
$$g(\alpha_1,\alpha_2)=g\left[(\alpha_1-\alpha_2)\left(\frac{\alpha_1-\alpha_2}{\alpha_1+\alpha_2}\right)^{0.2}\right].$$

{\it Conclusions} ---
We have shown that the known scaling laws for  $\alpha_c$
in the Heisenberg and the $XY$ model, i.e., 
Eqs.\ (\ref{scale1}) \& (\ref{scale0})
also hold for the conductance.
For the spin-rotationally invariant case we found a generalization for
acoustic phonons Eq.\ (\ref{scale2}) as well as a scaling relation for the gap
Eq.\ (\ref{scale2g}).
We also discussed the difference between two possible couplings to phonons
($\alpha_2=0$ \& $\alpha_2=-\alpha_1$). For the spin-rotationally invariant case 
the two ways of coupling are related by a simple scaling law Eq.\ 
(\ref{scale3}), whereas for the $XY$ chain the situation 
appears to be more complicated.
Assuming that our scaling laws are not restricted  
 we may make a prediction for the conductance of the  spin-Peierls model
Eq.\ (\ref{HI}) by putting together Eqs.\ (\ref{ItoII}), (\ref{scale2}) and (\ref{scale3}) as well as data
for the critical coupling from Ref.\ \onlinecite{Sandvik2}. We expect a step-like conductance which drops to zero when 
$\lambda=0.65\cdot2^{5/4}\omega_0\approx 1.55\omega_0$.

One final comment should be made.
Over the last few years, it has generally agreed upon
that the conductance measured in
experiments corresponds to the conductance of  a system coupled to 
noninteracting leads. 
The leads - which were not considered here - may strongly 
affect the conductance. \cite{Safi,MaslovStone}
We can argue however that the dimerization transition we observed
will drive the conductance to zero even if it occurs only 
locally, i.e. in the interacting part system and not in the leads.
Hence, we think that our results have a clear experimental consequence.

\bigskip
We would like to acknowledge useful discussions with D. Baeriswyl and 
financial support by the E.U. grant MIRG-CT-2004-510543.\vspace{-0.4cm}

\begin{appendix}
\section{MC update for the phonons}
\label{update}
We write the Hamiltonian Eq.\ (\ref{HII})
 in the form $H=\sum h_n^{(i,j,k)}$, where $n$ is the 
site (bond) index and indices $(i,j,k)$ denote different types of (local)
operators. 

The parameter $j$ (resp. $k$) assumes values from $-1$ to $1$
and refers to the change of the phonon number on site $n$ (resp. $n+1$)
resulting from the action of the operator $h_n^{(i,j,k)}$.
In the sequel we will refer to $j$ or $k$ as phonon parameters.
The index $i$ refers to the action on the spin degrees of freedom.
For example, the operators for different $i$ and $j=0=k$ read 
\begin{eqnarray*}h^{(1,0,0)}_n&=&JS^+_nS^-_{n+1}/2\qquad h^{(2,0,0)}=\left(h^{(1,0)}\right)^\dag \\
h^{(3,0,0)}&=&C+JS^z_nS^z_{n+1}
+\omega (a_n^\dag a_n^{\phantom{\dag}}+a_{n+1}^\dag 
a_{n+1}^{\phantom{\dag}}).\end{eqnarray*}

Following Ref.\ \onlinecite{Sandvik},
we expand the Boltzmann weight in terms of the local operators, i.e.,
we sample the operator strings
 $\prod_l^m \langle l| h(l)|l+1\rangle$ 
according to their weight in the
 Boltzmann factor ($|l\rangle$ are states in the 
$S^z$-phonon occupation-number basis and $h(l)$ is the 
 local operator $h_n^{(i,j,k)}$ which appears on the $l$th position of the string).
The spin degrees of freedom  can be  updated 
 using the very efficient scheme of operator loops. \cite{SandSyl} During this update the parameters $j$ and $k$ 
remain unchanged.
All that remains to be done is to 
devise an additional update for the phonon parameters.

This is achieved as follows: 
As the operators $a_n$ and $a_n^\dag$ appear always 
pairwise in the string, we have to increase the phonon parameter at
one position (of the string) and decrease it at another position. 
In principle the two positions can be chosen at random, but
here we want to point out  that this can be done more efficiently:
We choose an operator $h_n^{(i,j,k)}$ in the string. Next,
we decrease   (resp.\ increase, depending
on an appropriate probability) one of its phonon parameters
 (associated with a site which we will call $m$).
 For the previous (resp.\ next) operator in the string we choose---with
 a certain probability---one of three options: either (i) we update
the phonon number on  site  $m$ by $+1$ or (ii) we cancel the operation and restore the initial configuration, or (iii) we 
increase (resp.\ decrease) the phonon parameter  related with site $m$.
 In case (ii) or (iii) the procedure stops, while
in the first case it is iterated until we finally choose one of the other
 options.

 Even though one might suspect that
because of   option (ii)
 the algorithm is inefficient---since it is reminiscent of the 
bounce process in Ref.\ \onlinecite{SandSyl}---we find that
our new update scheme reduces the autocorrelation time
 significantly. The reason for the latter may be seen in the fact that 
we choose only one position of the two phonon operators at random, the other
  is---in some optimized way---chosen by the MC algorithm. 

\begin{figure}[h]
\epsfig{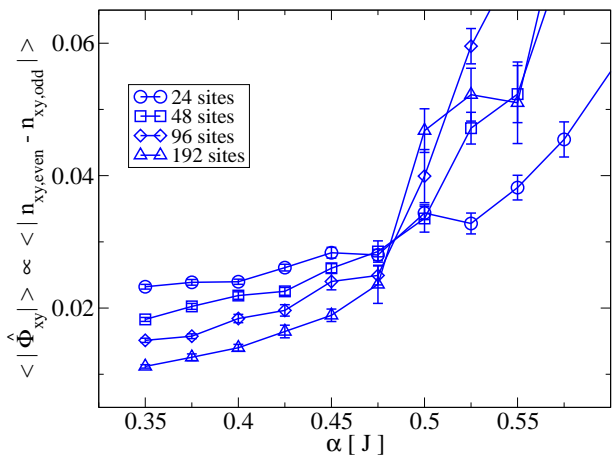}
\caption{Absolute value of the contributions of the staggered kinetic energy
$\langle |\hat \Phi_{xy}|\rangle$
(see text) for the $XY$-model with $\omega=J/8$ at $T=0.01J$. Here we use periodic boundary conditions and $10^4$ MC sweeps.
}
\label{gapJ0}
\end{figure}
\begin{figure}[h]
\epsfig{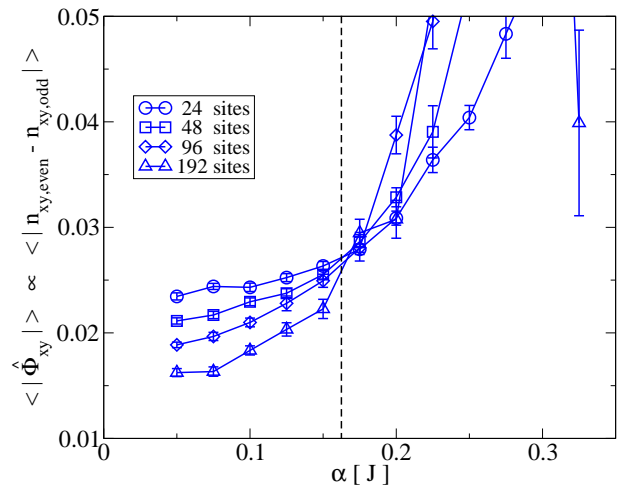}
\caption{Absolute value of the contributions of the staggered kinetic energy
$\langle |\hat \Phi_{xy}|\rangle$ (see text) for the Heisenberg-model with
 $\omega=J/8$ at $T=0.01J$. The dashed line give the critical coupling from Ref.\ \onlinecite{Sandvik2}. Here we use periodic boundary conditions and $10^4$ MC sweeps.
}
\label{gapJ1}
\end{figure}
\section{Evaluation of the critical coupling with QMC}
\label{critical}
 The simplest method to determine 
$\alpha_c$ with Monte Carlo is measuring the order parameter.
In our case we have two options for the order parameter: one 
 is $$\Phi_a=\sum_n(-1)^n(a_n+a_n^\dag)/N$$ the staggered bond length,\cite{Uhrig2} another is the 
staggered kinetic energy\cite{Sun} 
$$\Phi_{xy}=\sum_n(-1)^n(S_n^+S_{n+1}^-+S_n^-S_{n+1}^+)/(2N).$$
Since these operators are linear combinations of terms appearing in the
 Hamiltonian, there is a simple method to evaluate the expectation values with the SSE.\cite{Sandvik} To this end we define  $n_{a,n}$ 
(and $n_{xy,n}$)  which are the sum of the numbers of 
appearances of the operators $Ca_n$ and
 $Ca_n^\dag$ ($J/2S_n^+S_{n+1}^-$ and $J/2S_n^-S_{n+1}^+$) in the string of 
operators created and updated in the 
simulation process. The expectation values of the order parameters
 are related  with 
 the mean values of the numbers $n_{a,n}$ and $n_{xy,n}$;
 the relation being
 $$\langle \Phi_a\rangle=\sum_n(-1)^n\langle -n_{a,n}\rangle
/(N\beta C)$$
and $$\langle \Phi_{xy}\rangle=
\sum_n(-1)^n\langle -n_{xy,n}\rangle/(N\beta \tilde J).$$ 
Since the contributions to the order parameters may be positive or negative, it is 
difficult to detect a finite dimerization---especially,
 when we are close to the critical 
coupling. Hence, it is difficult  to decide whether we are in the gapped phase
 or not.  We therefore calculate the following expectation values
$$\langle |\hat\Phi_a|\rangle=
\Bigl\langle \Bigl|\sum_n(-1)^nn_{a,n}\Bigr|\Bigr\rangle/(N\beta C)$$
and $$\langle |\hat\Phi_{xy}|\rangle=
\Bigl\langle \Bigl|\sum_n(-1)^nn_{xy,n}\Bigr|\Bigr\rangle/(N\beta \tilde J).$$ 
(One should note that in general $\langle |\hat\Phi_a|\rangle\neq 
\langle |\Phi_a|\rangle$ where $|\Phi_a|=\sqrt{\Phi_a^2}$.)
These have non-zero values, but in the gapless phase they should go to zero 
in the thermodynamic limit.
We note that  while $\langle \Phi_a\rangle$ is independent of $C$
the expectation value $\langle |\hat\Phi_a|\rangle$ {\it does} depend on the shift 
$C$. Because of this and the fact that $\langle |\hat\Phi_a|\rangle>\langle |\hat\Phi_{xy}|\rangle$ (which suggests that $\langle |\hat\Phi_{xy}|\rangle$ decays faster with $N$) we prefer the second order parameter to the first and will only discuss $\langle |\hat\Phi_{xy}|\rangle$.

 Results for $\langle |\hat\Phi_{xy}|\rangle$,
 the staggered kinetic energy, are 
presented in  Figs.\ \ref{gapJ0} and \ref{gapJ1}. For the $XY$
model (Fig.\ \ref{gapJ0})   the order parameter
$\langle |\hat\Phi_{xy}|\rangle$
 decreases with system size as long as $\alpha_1\leq 0.475J$. For $\alpha_1\geq 0.5J$ it seems 
to increase with system size rather than decrease. We conclude  that 
$\alpha_c<0.5J$.  
A correct assessment of the critical coupling remains difficult.
A rough estimate for $\alpha_c$ is obtained by assuming that
 $\alpha_c$ lies in the interval $0.475J<\alpha_1<0.5J$
 where the  behavior of 
$\langle |\hat\Phi_{xy}|\rangle$
 changes. This yields $\alpha_c\approx 0.4875J\pm 0.0125J$.
In principle, this estimate  gives only an upper bound for $\alpha_c$,
but when we apply the same method to  
 the spin-rotationally invariant case,  we find good agreement
 with the value for $\alpha_c$
 from Ref.\ \onlinecite{Sandvik2} (see Fig.\ \ref{gapJ1}).

\section{Mean Field}
\label{MeanField}
To better understand the physics of the 
model a mean field analysis is helpful. The main idea is to replace
the displacements $(a_n+a_n^\dag)/\sqrt{2}$ by classical variables $x_n$.
A common but general ansatz for these variables is 
$x_n=\bar x+(-1)^n\Delta$.
Substituting this ansatz into the Hamiltonian Eq.\ (\ref{HII}) we obtain,

\begin{eqnarray}
H_{MF}&=&\sum_n J_n S_n\cdot S_{n+1}\\
J_n=J&+&\sqrt{2}(\alpha_1+\alpha_2)\bar x+
(-1)^n \sqrt{2}(\alpha_1-\alpha_2)\Delta\nonumber.
\label{HSCA}
\end{eqnarray}

One sees that the spin-Peierls model Eq.\ (\ref{HI})
 does not depend on $\bar x$ because of $\alpha_1=-\alpha_2$.
At the mean field theory level we arrive at two conclusions,
(i) there is a shift in $J$ proportional to $(\alpha_1+\alpha_2)$
and (ii) there is a dimerization transition where the dimerization is 
proportional to $(\alpha_1-\alpha_2)$.
In the spin-Peierls model ($\alpha_1=-\alpha_2$) the shift in $J$
is absent, whereas in a toy model with $\alpha_1=\alpha_2$ no dimerization 
should be found.
In our simulations ($\alpha_1,\alpha_2\geq 0$) we always
have a shift but, since the conductance at $T=0$ does not depend 
on $J$, we expect that $g$ is independent of $\alpha_1+\alpha_2$ and we 
arrive at the scaling law 
\begin{equation} g(\alpha_1,\alpha_2)=g(\alpha_1-\alpha_2).
\label{scale3a}
\end{equation}

\end{appendix}


\end{document}